\documentclass[12pt]{article}

\usepackage{amsmath,amssymb}
\usepackage{booktabs, threeparttable, multirow}
\usepackage{graphicx}
\usepackage[round]{natbib}
\usepackage{authblk}
\usepackage{color}
\usepackage{geometry}
\usepackage{pdfpages}

\def\T{{\mathrm{\scriptscriptstyle T} }}
\newcommand\independent{\protect\mathpalette{\protect\independenT}{\perp}}
\def\independenT#1#2{\mathrel{\rlap{$#1#2$}\mkern2mu{#1#2}}}
\DeclareMathOperator{\var}{var}
\DeclareMathOperator{\cov}{cov}
\DeclareMathOperator{\diag}{diag}
\DeclareMathOperator{\logit}{logit}

\allowdisplaybreaks[4]
\newtheorem{theorem}{Theorem}

\newtheorem{assumption}{Assumption}

\newtheorem{remark}{Remark}
\newtheorem{example}{Example}
\title{\bf Analysis of Broken Randomized Experiments by Principal Stratification}

\author[1]{Qinqing Liu}
\author[1]{Xiang Peng$^{\dagger}$}
\author[1]{Tao Zhang$^{\dagger}$\thanks{Xiang Peng and Tao Zhang are both corresponding authors.}}
\author[2]{Yuhao Deng}
\affil[1]{\small ~Business School, Soochow University, Suzhou, Jiangsu, China}
\affil[2]{\small ~Department of Biostatistics, School of Public Health, University of Michigan, Ann Arbor, Michigan, USA}

\date{}

\begin{document}

\maketitle

\vspace{-24pt}

\begin{abstract}
Although randomized controlled trials have long been regarded as the ``gold standard'' for evaluating treatment effects, there is no natural prevention from post-treatment events. For example, non-compliance makes the actual treatment different from the assigned treatment, truncation-by-death renders the outcome undefined or ill-defined, and missingness prevents the outcomes from being measured. In this paper, we develop a statistical analysis framework using principal stratification to investigate the treatment effect in broken randomized experiments. The average treatment effect in compliers and always-survivors is adopted as the target causal estimand. We establish the asymptotic property for the estimator. To relax the identification assumptions, we also propose an interventionist estimand defined in compliers by adjusting for baseline covariates. We apply the framework to study the effect of training on earnings in the Job Corps study and find that the training program improves employment and earnings in the long term. \par
\end{abstract}

\section{Introduction} \label{sec1}

\subsection{Broken randomized experiments}

Randomized controlled trials (RCTs) have long been regarded as the ``gold standard'' for evaluating treatment effects because randomization leads to an identical distribution of baseline characteristics in the treated group and control group. Measured and unmeasured factors that affect the outcome are naturally balanced. As a result, the difference in average outcomes between the treated and control group must result from the treatment. However, although the effect of pre-treatment factors is eliminated using the randomized design, there is no guarantee that post-treatment events that affect the primary outcome can be prevented. Some individuals have a higher probability of complying with the treatment assignment or have a higher probability of observing the outcome. The distributions of baseline characteristics would be different if we only compare the individuals with measured and well-defined outcomes.

Some post-treatment events lead to broken randomized experiments \citep{rubin1976inference, barnard1998broader, hirano2000assessing, rubin2006causal, mcconnell2008truncation}. The most typical post-treatment events are non-compliance, truncation-by-death, and missingness.
\begin{enumerate}
\item Non-compliance. The individuals do not adhere to the treatment assignments. For example, when studying the effect of a new drug on a disease, some patients assigned to receive the treatment were allergic to the drug. Practitioners offer these patients a placebo even though the randomization programme assigns these patients to the drug. These patients in the assigned treated group actually received control. The distributions of baseline characteristics are different between the actually treated individuals and actually controlled individuals because there are more allergic individuals in the actually controlled group \citep{frangakis1999addressing}.
\item Truncation-by-death. The outcomes for some patients are not well-defined. For example, to study the effect of a surgery on the quality of life, some patients die before the end of the study. It is not reasonable to record the quality of life as any value because we can not define the quality of life for dead people. The observed survivors between the treated group and control group have different characteristics if the mortality risks are different between the groups. A popular terminology is the ``survivor bias'' \citep{zhou2005survival, snyder2009relationship}. Another example is rescue medication, in which patients receive additional medication in addition to the assigned treatment if emergencies occur. The outcome would be a result of the joint effect of the initial treatment and rescue medication. To evaluate the pure effect of the initial treatment, we need to separate the effect of rescue medication from the total effect.
\item Missingness. The outcome of interest is well-defined but is not collected. Missingness is different from truncation-by-death. For example, a patient quits the experiment after moving to another city. Although this patient has an outcome (quality of life), practitioners do not record the outcome. The probability of missingness can depend on individual characteristics.
\end{enumerate}

Non-compliance, truncation by death, and missingness can happen together in an experiment. The following gives two concrete examples to illustrate the broken randomized experiments.

\begin{example}[Natural experiments]
Suppose the research interest is the return to an additional year of schooling. In economics, the season when a person was born is used as an instrumental variable. Those born late in the year need to wait longer to attend school, so they have a higher probability of entering the job market with a shorter education when they celebrate their adult birthday. It seems that the birth season is randomized by nature. The birth season is the assigned treatment, and whether completing school is the actual treatment. Some individuals would always drop out regardless of birth season, while some would always complete high school, resulting in non-compliance. Another problem is that if people are not employed, their earnings cannot be defined, resulting in truncation-by-death. Those staying at home may have excellent potential to earn money, but they decide not to work because the salary offered by the employer does not meet their expectation. It is inappropriate to record his wage as zero.
\end{example}

\begin{example}[Encouragement designs]
Suppose the research interest is the effect of a vaccine on virus load after infection. Subject to ethical issues, practitioners can not force patients to take vaccines or not to take vaccines. Therefore, encouragement designs are usually adopted, where patients are randomly encouraged to take the vaccine. With encouragement, patients are more likely to take the vaccine. Those who are not encouraged can actually take the vaccine, and those who are encouraged can reject the vaccine by their own judgment, resulting in non-compliance. On the other hand, if a patient is not infected due to cautious self-protection, the virus load cannot be defined, resulting in truncation by death. The vaccine may have no effect on infection, but it has a significant effect on reducing the virus load. We need measurements of virus load in the target population to compare treatments, so the target population should not include the uninfected people.
\end{example}

Our work is motivated by the Job Corps (JC) education and job training program for disadvantaged youths \citep{schochet2001national, burghardt2001does}. In the Job Corps study, the disadvantaged youths are randomly provided with an opportunity to take a job training program, which lasts for several years. Employment status and weekly earnings were recorded. A lot of work has been done to study the effect of this job training program \citep{schochet2008does, zhang2009likelihood, flores2010learning, flores2012estimating, chen2015bounds}. Nevertheless, the simultaneous existence of non-compliance and truncation-by-death provides great challenges to evaluating the pure effect of the program. Statistically sound and interpretable methods are still desired to analyze broken randomized experiments.

\subsection{Principal stratification}

A lot of methodologies have been proposed to deal with non-compliance. The most intuitive idea is to adopt the intention-to-treat strategy to study the effect of the treatment assignment \citep{ten2008intent, mcnamee2009intention, gupta2011intention}. The treatment effect is a joint effect of the treatment assignment and the following post-treatment events. The as-treated and per-protocol strategies focus on some subpopulations defined by observed compliance types. However, these estimands do not necessarily have meaningful causal interpretations because the target population is determined by observed post-treatment variables, which are essentially realizations of potential outcomes. The people receiving treatment and receiving control under comparison are different. The instrumental variable approach regards the treatment assignment as an instrumental variable for the actually received treatment \citep{angrist1996identification, martens2006instrumental, tan2006regression, baiocchi2014instrumental}. A homogeneity or non-interaction assumption is required to identify the treatment effect in the overall population \citep{klein2010heterogeneous, wang2018bounded}.

The instrumental variable approach is closely related to principal stratification \citep{frangakis2002principal, lipkovich2022using}. With non-compliance, the target population only consists of compliers, who will always comply with the assigned treatment regardless of being assigned active treatment or control. Under the potential outcomes framework \citep{rubin1974estimating}, the target population is determined by joint potential outcomes, so identifying the complier treatment effect requires untestable assumptions. Usually, monotonicity is assumed for point-identification, saying that there are no defiers \citep{chetverikov2017nonparametric, swanson2018challenging}. Some alternatives include relaxing monotonicity to stochastic monotonicity \citep{small2017instrumental}, or considering bounds for the treatment effect \citep{santos2012inference, clarke2012instrumental, swanson2018partial, lange2020iv, gabriel2023nonparametric, gabriel2025impact}.

With truncation-by-death, a principal stratum of always-survivors is adopted as the target population, among which both potential outcomes are well defined \citep{ding2011identifiability}. In economic studies with earnings as the primary outcome and employment as the intermediate variable, this principal stratum is the always-employed \citep{zhang2003estimation, zhang2009likelihood}. A monotonicity assumption on survival and a principal ignorability assumption (also referred to as explainable nonrandom survival) are usually made to identify the principal stratum treatment effect \citep{jo2009use, stuart2015assessing, feller2017principal, jiang2022multiply, qu2023assessing}. Alternatives to principal ignorability include constructing bounds \citep{imai2008sharp, long2013sharpening, yang2016using}, imposing selection models \citep{shepherd2006sensitivity, jemiai2007semiparametric} and invoking substitutional variables \citep{tchetgen2014identification, wang2017identification}.

Of note, truncation-by-death is a different problem from missingness. The former means that the outcome is not well-defined, whereas the latter means that the outcome exists but is not observed. To deal with missing data, the selection model and the pattern mixture model assume different data generating mechanisms and adopt different estimands \citep{little1993pattern, little2008selection, little2017conditions}. Although missingness may depend on individual characteristics, we simply consider the missingness mechanism as a nuisance in this article. We do not incorporate the pattern of missingness in the principal strata by assuming missingness is random.

Principal stratification can be used to deal with multiple post-treatment events. \citet{mattei2007application} extended the principal stratification framework to randomized experiments suffering from treatment non-compliance, missing outcomes, and truncation by death. In their work, there are only two types of compliance behaviors, namely compliers and never-takers. They developed a Bayesian model to estimate the causal parameters. \citet{frumento2012evaluating} assumed a mixture distribution for potential outcomes and identified the parameters using the EM algorithm. As a weakness of mixture models, identification is fragile when the outcome behaviors are similar between principal strata. \citet{chen2015bounds} constructed a bound for the average treatment effect in a principal stratum. The validity of bounds relies on other assumptions and can be wide, so it may be inconvenient to make inferences based on the bounds. There is a lack of theoretical properties on the estimators, which hinders statistical inference about related scientific questions.

\subsection{Contribution and the article structure}

Few literature have studied the (nonparametric) identification and estimation of treatment effects with non-compliance, truncation-by-death, and missingness simultaneously. We fill this gap by developing a statistical analysis framework in broken randomized experiments using principal stratification. We define the causal estimand on survived compliers, who will always comply with the assigned treatment and survive regardless of treatments. Under some assumptions, we show the identifiability of the estimand, and establish the asymptotic properties. To relax the assumptions, we propose an interventionist estimand defined on compliers by adjusting for baseline covariates. We contribute to empirical economics by analyzing the Job Corps data. We find that the training program has a positive effect on emplyment and earnings in the long term.

The remainder of this article is organized as follows. Section \ref{sec2} introduces the notations and assumptions for identification. We provide the identification results. Section \ref{sec3} introduces the estimation and inference procedure. We establish the asymptotic property for the estimator of the principal stratum average treatment effect. In Section \ref{sec4}, we conduct simulation studies to assess the performance of the treatment effect estimator and the coverage of the confidence interval obtained from the asymptotic formula. In Section \ref{sec5}, the proposed framework is applied to study the effect of the Job Corps. Finally, we point out some limitations and research directions in Section \ref{sec6}. Extensions of the proposed framework are provided in the Supplementary Material.

\section{Framework, estimand and identification} \label{sec2}

\subsection{Notations and assumptions}

Let $Z$ be the binary treatment assignment, where $Z=1$ denotes the active treatment and $Z=0$ denotes the control. Following the potential outcomes framework, let $D(z)$ be the binary treatment an individual actually received if assigned to the treatment arm $z$. Let $S(z,d)$ be the binary indicator of whether the outcome of interest is defined. In the original truncation-by-death context, $S(z,d)=1$ means the individual survives, and $S(z,d)=0$ means the individual dies under the treatment assignment $z$ and actual treatment $d$. In the employment context, $S(z,d)=1$ means employed, and $S(z,d)=0$ means unemployed. Let $Y(z,d)$ be the potential outcome of interest, which is only well defined when $S(z,d)=1$. We supplementarily denote $Y(z,d)=*$ (a non-real number) if $S(z,d)=0$. 

We make the stable unit treatment value assumption (SUTVA). That is, there is no interference between individuals, and there is only one version of active treatment. Let $D$, $S$, and $Y$ be the observed treatment, survival status, and outcome, respectively. We assume consistency to link potential outcomes with observed outcomes.

\begin{assumption}[Consistency] \label{ass:con}
$D(Z)=D$, $S(Z,D)=S$, $Y(Z,D)=Y$.
\end{assumption}

Consistency implies that the potential variables $S(z,d)$, $Y(z,d)$ can be observed if $Z=z$ and $D=d$ in the experiment. Let $\Delta_S$ be the missingness indicator of the survival variable and $\Delta_Y$ be the missingness indicator of the primary outcome, with 1 indicating observation and 0 indicating missingness. Suppose the sample includes $n$ independent individuals randomly drawn from a super-population. We use the subscript $i$ to denote the measurements for the $i$th individual when necessary. The observed sample is $\mathcal{O} = \{O_i = (Z_i, D_i, \Delta_{Si}, S_i\Delta_{Si}, \Delta_{Yi}, Y_i\Delta_{Yi}): i=1,\ldots,n\}$.

We make the following assumptions, which are common in literature on instrumental variables (non-compliance).

\begin{assumption}[Randomization] \label{ass:ran}
$Z \independent (D(z),S(z,d),Y(z,d): z,d=0,1)$.
\end{assumption}
\begin{assumption}[Monotonicity on compliance] \label{ass:mon}
$D(1) \ge D(0)$ almost surely.
\end{assumption}
\begin{assumption}[Exclusion restriction] \label{ass:er}
$S(1,d)=S(0,d)$ for $d=0,1$. $Y(1,d)=Y(0,d)$ if $S(1,d)=S(0,d)=1$ for $d=0,1$.
\end{assumption}
\begin{assumption}[Substitution relevance] \label{ass:sr}
$E\{D(1)\} \neq E\{D(0)\}$.
\end{assumption}

\begin{figure}
\centering
\includegraphics[width=0.4\textwidth]{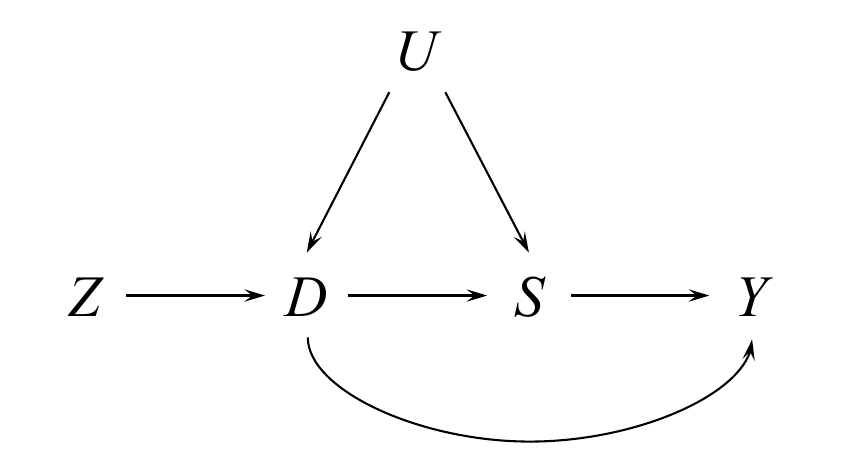}
\caption{A directed acyclic graph (DAG) of observed variables.} \label{fig:dag}
\end{figure}

Assumption \ref{ass:ran} states that the treatment assignment is completely randomized. There should be no measured and unmeasured confounding. Therefore, the treatment assignment is independent of all potential variables. In stratified randomized experiments, there may be baseline covariates. For identification, the analysis should be conducted in each level of covariates. Assumption \ref{ass:mon} rules out defiers. The treatment assignment (encouragement) should have a non-negative effect on the actual treatment for each individual. Assumption \ref{ass:er} states that the treatment assignment can only exert its effect on survival or outcome through the actual treatment. There is no direct effect of the treatment assignment on the survival or outcome. Under Assumption \ref{ass:er}, we can write $S(z,d)=S(d)$ and $Y(z,d)=Y(d)$. Assumption \ref{ass:sr} ensures that compliers, who would always adhere to the treatment assignment, must exist. Figure \ref{fig:dag} shows the directed acyclic graph (DAG) of observed variables. The unmeasured confounder $U$ can affect $D$ and $S$ but should not affect $Z$.

Stratified by the potential values of $(D(1), D(0))$ and $(S(1), S(0))$, the whole population can be divided into eight principal strata, as listed in Table \ref{tab1}. The treatment effect is defined as the contrast of $Y(1)$ and $Y(0)$ on the target population. However, there is only one principal stratum, the survived compliers ($cl$), in which both $Y(1)$ and $Y(0)$ are well defined. In other principal strata, at least one potential outcome $Y(z)$ ($z=0,1$) is undefined. Therefore, we should restrict the target population to the survived compliers. We define the principal average causal effect (PACE) in the survived compliers as
\[
\tau = E\{Y(1)-Y(0)\mid G=cl\},
\]
which measures the magnitude of average treatment effects in compliers who would always survive regardless of the treatment received. In the Job Corps study, a survived complier must satisfy two conditions. First, he/she will join the training if assigned to the treated group, and will not join the training if assigned to the control group. Second, no matter he/she joins the program or not, he/she will always be employed.

\begin{table}
\centering
\caption{Principal strata in the population. The notation \checkmark is used to denote well-definedness and $\times$ to denote undefinedness} \label{tab1}
\begin{tabular}{cccccccl}
\hline
Stratum & $D(1)$ & $D(0)$ & $S(1)$ & $S(0)$ & $Y(1)$ & $Y(0)$ & Description \\ \hline
$al$ & 1 & 1 & 1 & $\times$ & \checkmark & $\times$ & Survived always-takers \\
$ad$ & 1 & 1 & 0 & $\times$ & $\times$ & $\times$ & Dead always-takers \\
$cl$ & 1 & 0 & 1 & 1 & \checkmark & \checkmark & Survived compliers \\
$cp$ & 1 & 0 & 1 & 0 & \checkmark & $\times$ & Protected compliers \\
$ch$ & 1 & 0 & 0 & 1 & $\times$ & \checkmark & Harmed compliers \\
$cd$ & 1 & 0 & 0 & 0 & $\times$ & $\times$ & Doomed compliers \\
$nl$ & 0 & 0 & $\times$ & 1 & $\times$ & \checkmark & Survived never-takers \\
$nd$ & 0 & 0 & $\times$ & 0 & $\times$ & $\times$ & Dead never-takers \\ \hline
\end{tabular}
\end{table}

Some outcomes are truncated by death and subject to missingness. We assume that $Y$ will be missing if $S$ is missing. The assumptions below are also required for identification to account for truncation by death (the problem of partially defined outcomes) and unintended missingness.

\begin{assumption}[Principal ignorability] \label{ass:pi}
$E\{Y(1)\mid G=cl\} = E\{Y(1)\mid G=cp\}$ and $E\{Y(0)\mid G=cl\} = E\{Y(0)\mid G=ch\}$.
\end{assumption}
\begin{assumption}[Missingness at random] \label{ass:mar}
$\Delta_S \independent S(d) \mid Z=z, D(z)=d$, and $\Delta_Y \independent Y(d) \mid Z=z, D(z)=d, S(d)=1$, for $z,d=0,1$.
\end{assumption}

Assumption \ref{ass:pi} means that the potential outcome of interest $Y(d)$ does not have cross-world reliance on $S(1-d)$ as long as $S(d)=1$ \citep{jo2009use}. This assumption is sometimes referred to as explainable nonrandom survival \citep{hayden2005estimator, egleston2007causal, egleston2009estimation} to indicate that the survival cannot causally predict the potential outcome of interest. For notation simplicity, denote the compliance strata
\begin{align*}
a &= \{D(1)=D(0)=1\} = \{al, ad\}, \\
c &= \{D(1)>D(0)\} = \{cl, cp, ch, cd\}, \\
n &= \{D(1)=D(0)=0\} = \{nl, nd\}.
\end{align*}
Let $p_{g} = P(G=g)$ be the proportion of principal stratum $g$.
Assumption \ref{ass:pi} merges always-survivors and protected subpopulations in the treated group ($cl$ and $cp$), and merges always-survivors and harmed populations in the control group ($cl$ and $ch$) for compliers. Under Assumption \ref{ass:pi},
\begin{align*}
E\{Y(1)\mid G=cl\} &= E\{Y(1)\mid G=cp\} = E\{Y(1)\mid S(1)=1,c\}, \\
E\{Y(0)\mid G=cl\} &= E\{Y(0)\mid G=ch\} = E\{Y(0)\mid S(0)=1,c\}, \\
E\{Y(1)\mid G=al\} &= E\{Y(1)\mid S(1)=1, a\}, \\
E\{Y(0)\mid G=nl\} &= E\{Y(0)\mid S(0)=1, n\}.
\end{align*}

Assumption \ref{ass:mar} means that the missingness of an variable is independent of potential outcomes given observed data before this variable. Although $Y(d)$ is not always well-defined, the product $Y(d)S(d)$ is always well-defined. Under Assumptions \ref{ass:con}, \ref{ass:pi}, and \ref{ass:mar}, we have
\begin{align*}
E\{Y(d)S(d)\mid Z=z, D=d\} &= E\{S\mid Z=z, D=d, \Delta_S=1\} \\
&\quad\cdot E\{Y\mid Z=z, D=d, S=1, \Delta_Y=1\}
\end{align*}
for $z,d=0,1$. Therefore, we can identify the mean product of the survival and outcome in any subgroup stratified by the assigned treatment and received treatment. This equation allows us to estimate the conditional expectations of $S$ and $Y$ separately using available data, improving the efficiency of estimation.

\subsection{Identification of the PACE}

There are three levels of data generation: compliance types, survival types, and outcome. The first-stage regression gives the proportions of compliance groups. The individuals receiving treatment come from always-takers and compliers in the treated group, and the individuals receiving treatment come solely from always-takers in the control group. Therefore, we have
\begin{align*}
E\{D\mid Z=1\} &= p_{a} + p_{c}, \\
E\{D\mid Z=0\} &= p_{a},
\end{align*}
from which we can solve
\begin{align*}
p_{a} &= E\{D\mid Z=0\}, \\
p_{c} &= E\{D\mid Z=1\}-E\{D\mid Z=0\}, \\
p_{n} &= 1-E\{D\mid Z=1\}.
\end{align*}

The second-stage regression gives the proportions of survival groups in compliers. Solving the following equations,
\begin{align*}
E\{S\mid Z=1,D=1\} &= \frac{p_a}{p_a+p_c}P(S(1)=1\mid a) + \frac{p_c}{p_a+p_c}P(S(1)=1\mid c), \\
E\{S\mid Z=1, D=0\} &= P(S(0)=1\mid n), \\
E\{S\mid Z=0, D=1\} &= P(S(1)=1\mid a), \\
E\{S\mid Z=0, D=0\} &= \frac{p_c}{p_c+p_n}P(S(0)=1\mid c) + \frac{p_n}{p_c+p_n}P(S(0)=1\mid n),
\end{align*}
we obtain
\begin{align*}
P(S(1)=1\mid c) &= \frac{\{p_c+p_a\}E\{S\mid Z=1,D=1\}-p_aE\{S\mid Z=0,D=1\}}{p_c}, \\
P(S(0)=1\mid c) &= \frac{\{p_c+p_n\}E\{S\mid Z=0,D=0\}-p_nE\{S\mid Z=1,D=0\}}{p_c}.
\end{align*}

The third-stage regression links the outcomes with principal strata. Note that $YS$ is always well-defined, 
\begin{align*}
E\{YS\mid Z=1,D=1\} &= \frac{p_a}{p_a+p_c}P(S(1)=1\mid a)E\{Y(1)\mid S(1)=1, a\} \\
&\quad +\frac{p_c}{p_a+p_c}P(S(1)=1\mid c)E\{Y(1)\mid S(1)=1, c\}, \\
E\{YS\mid Z=1,D=0\} &= P(S(0)=1\mid n)E\{Y(0)\mid S(0)=1, n\}, \\
E\{YS\mid Z=0,D=1\} &= P(S(1)=1\mid a)E\{Y(1)\mid S(1)=1, a\}, \\
E\{YS\mid Z=0,D=0\} &= \frac{p_c}{p_c+p_n}P(S(0)=1\mid c)E\{Y(0)\mid S(0)=1, c\} \\
&\quad +\frac{p_n}{p_c+p_n}P(S(0)=1\mid n)E\{Y(0)\mid S(0)=1, n\}.
\end{align*}
The solution of the above equations is
\begin{align*}
E\{Y(1)\mid G=cl\} &= \frac{\{p_a+p_c\}E\{YS\mid Z=1,D=1\}-p_aE\{YS\mid Z=0,D=1\}}{p_c P(S(1)=1\mid c)} \\
&= \frac{\left[\begin{aligned}
&E\{DS\mid Z\Delta_S=1\}E\{Y\mid ZDS\Delta_Y=1\} \\[-4pt]
&\quad -E\{DS\mid (1-Z)\Delta_S=1\}E\{Y\mid (1-Z)DS\Delta_Y=1\}
\end{aligned}\right]
}{E\{DS\mid Z\Delta_S=1\}-E\{DS\mid (1-Z)\Delta_S=1\}}, \\
E\{Y(0)\mid G=cl\} &= \frac{\{p_c+p_n\}E\{YS\mid Z=0,D=0\}-p_nE\{YS\mid Z=1,D=0\}}{p_c P(S(0)=1\mid c)} \\
&= \frac{\left[\begin{aligned}
&E\{(1-D)S\mid Z\Delta_S=1\}E\{Y\mid Z(1-D)S\Delta_Y=1\} \\[-4pt] 
&\quad -E\{(1-D)S\mid (1-Z)\Delta_S=1\}E\{Y\mid (1-Z)(1-D)S\Delta_Y=1\}\end{aligned}\right]
}{E\{(1-D)S\mid Z\Delta_S=1\}-E\{(1-D)S\mid (1-Z)\Delta_S=1\}}.
\end{align*}

Afterward, the identifiability of PACE follows immediately.

\begin{theorem} \label{thm1}
Under Assumptions \ref{ass:con}--\ref{ass:mar}, the expectation of potential outcomes in survived compliers $E\{Y(d) \mid cl\}$ and the principal average causal effect $\tau$ are nonparametrically identifiable.
\end{theorem}

In particular, we have the following results as corollaries of Theorem \ref{thm1} when the identification assumptions hold.

\begin{remark} \label{rm1}
If there is no truncation-by-death, i.e., $P(S(0)=1\mid D(1),D(0))=1$, then the principal average causal effect is reduced to the Wald estimator in instrumental variables,
\begin{align*}
\tau &= \frac{E\{Y\mid Z=1,\Delta_Y=1\}-E\{Y\mid Z=0,\Delta_Y=1\}}{E\{D\mid Z=1\}-E\{D\mid Z=0\}}.
\end{align*}
\end{remark}
\begin{remark} \label{rm2}
If there is no non-compliance, i.e., $P(D(1)>D(0))=1$, then the principal average causal effect is reduced to the observed survivor-case contrast,
\begin{align*}
\tau=E\{Y\mid Z=1,S=1,\Delta_Y=1\}-E\{Y\mid Z=0,S=1,\Delta_Y=1\}.
\end{align*}
In some circumstances, the principal ignorability (Assumption \ref{ass:pi}) may seem strong. However, the principal average causal effect would not be identifiable without introducing additional information on the cross-world reliance of $Y$ on $S$ or invoking a substitutional variable for the principal stratum $G$ if the principal ignorability assumption is violated.
\end{remark}
\begin{remark} \label{rm3}
If there is no missingness, then the principal average causal effect is
\begin{align*}
\tau &= \frac{E\{DSY\mid Z=1\}-E\{DSY\mid Z=0\}}{E\{DS\mid Z=1\}-E\{DS\mid Z=0\}} \\
&\quad -\frac{E\{(1-D)SY\mid Z=1\}-E\{(1-D)SY\mid Z=0\}}{E\{(1-D)S\mid Z=1\}-E\{(1-D)S\mid Z=0\}}.
\end{align*}
\end{remark}

In the presence of missingness, we can use standard multiple imputation methods to impute the missing values and use the formula in Remark \ref{rm3} to estimate the principal average causal effect \citep{harel2007multiple, kenward2007multiple, rubin2018multiple}. 

It is noted that the proportions of principal strata $cl$, $cp$, $ch$, and $cd$ defined by the compliance type and survival type are not identifiable without additional assumptions. To identify the proportion of survived compliers, a typical assumption is the individual-level monotonicity on $S$, i.e., $S(1) \ge S(0)$ almost surely, under which
\[
P(G=cl)=p_{c}P(S(0)=1\mid c)=E\{(1-D)S\mid Z=0\}-E\{(1-D)S\mid Z=1\}.
\]
Whether this assumption is plausible depends on specific scientific questions. For example, this monotonicity does not hold for the Job Corps data because joining the training may raise psychological expectations about wages \citep{frumento2012evaluating}. Still, it may be reasonable in some medical applications where the active treatment is beneficial for survival \citep{ding2011identifiability, wang2017identification}. The non-identifiability of principal strata proportions brings difficulty aggregating the conditional causal effect at each covariate level to the average causal effect in the overall population. More details about identification are provided in the Supplementary Material.

\section{Estimation and inference} \label{sec3}

Consider a completely randomized experiment of size $n$ with a possibly unknown treatment probability $P(Z=1)=1-P(Z=0)=r$. Suppose there is no missingness; otherwise, one can use multiple imputations or other methods to generate a completed dataset first. The asymptotic variance of the estimator can be calculated by Rubin's rule from completed datasets \citep{rubin1986multiple, white2011multiple, rubin2018multiple}.

Let $n_{zds}=\sum_{i=1}^{n}1(Z_i=z,D_i=d,S_i=s)$ for $z,d,s=0,1$, and use dot ``$\cdot$'' to represent summation with respect to its/their corresponding argument(s). Let $\mu_d=E\{Y(d)\mid G=cl\}$ for $d=0,1$. Denote 
\begin{align*}
&\theta_1 = E\{D\mid Z=1\}, \ \theta_{11} = E\{S\mid ZD=1\}, \ \theta_{10} = E\{S\mid Z(1-D)=1\}, \\
&\theta_0 = E\{D\mid Z=0\}, \ \theta_{01} = E\{S\mid (1-Z)D=1\}, \ \theta_{00} = E\{S\mid (1-Z)(1-D)=1\},
\end{align*}
and let $f(y;\theta_{zd1})$ be the density function of $Y$ given $Z=z$, $D=d$ and $S=1$ with mean $b(\theta_{zd1})=E\{Y\mid Z=z,D=d,S=1\}$. Suppose there are no common parameters in $\theta_{zd1}$ for $z,d=0,1$. Write the unknown parameter as
\[
\theta=(r,\theta_1,\theta_0,\theta_{11},\theta_{10},\theta_{01},\theta_{00},\theta_{111},\theta_{101},\theta_{011},\theta_{001})^{\T}.
\]
The likelihood for $\theta$ is
\begin{align*}
L(\theta;\mathcal{O}) &= (\theta_1)^{n_{11\cdot}}(1-\theta_1)^{n_{10\cdot}} \cdot (\theta_{11})^{n_{111}}(1-\theta_{11})^{n_{110}} \cdot (\theta_{10})^{n_{101}}(1-\theta_{10})^{n_{100}} \\
&\quad \cdot (\theta_0)^{n_{01\cdot}}(1-\theta_0)^{n_{00\cdot}} \cdot (\theta_{01})^{n_{011}}(1-\theta_{01})^{n_{010}} \cdot (\theta_{00})^{n_{001}}(1-\theta_{00})^{n_{000}} \\
&\quad \cdot \prod_{i:S_i=1}f(y_i;\theta_{Z_iD_i1}) \cdot r^{n_{1\cdot\cdot}}(1-r)^{n_{0\cdot\cdot}}.
\end{align*}
The maximized likelihood estimate of $\theta$ is given by
\[
\widehat\theta = \left(\frac{n_{1\cdot\cdot}}{n}, \frac{n_{11\cdot}}{n_{1\cdot\cdot}}, \frac{n_{01\cdot}}{n_{0\cdot\cdot}}, \frac{n_{111}}{n_{11\cdot}}, \frac{n_{101}}{n_{10\cdot}}, \frac{n_{011}}{n_{01\cdot}}, \frac{n_{001}}{n_{00\cdot}}, \widehat\theta_{111}, \widehat\theta_{101}, \widehat\theta_{011}, \widehat\theta_{001}\right)^{\T},
\]
where $\widehat\theta_{zd1}$ solves $\sum_{Z_i=z,D_i=d,S_i=1}\partial\log f(y_i;\theta_{zd1})/\partial\theta_{zd1}=0$, with
\begin{align*}
\cov(\widehat\theta) &= \diag\bigg\{\frac{r(1-r)}{n}, \frac{\theta_{1}(1-\theta_{1})}{nr}, \frac{\theta_{0}(1-\theta_{0})}{n(1-r)}, \frac{\theta_{11}(1-\theta_{11})}{nr\theta_1}, \frac{\theta_{10}(1-\theta_{10})}{nr(1-\theta_1)}, \\
&\qquad\qquad \frac{\theta_{01}(1-\theta_{01})}{n(1-r)\theta_0}, \frac{\theta_{00}(1-\theta_{00})}{n(1-r)(1-\theta_0)}, \var(\widehat\theta_{111}), \var(\widehat\theta_{101}), \var(\widehat\theta_{011}), \var(\widehat\theta_{001})\bigg\}.
\end{align*}
If we use the sample average $\widehat\theta_{zd1}$ of the outcomes in $\{Z=z,D=d,S=1\}$ to estimate the conditional mean outcome $b(\theta_{zd1})$, then $\var(\widehat\theta_{zd1})$ is the variance of the observed outcomes in $\{Z=z,D=d,S=1\}$ divided by the group size $n_{zd1}$.

Since we can express the mean potential outcome $\mu_d = E\{Y(d) \mid G=cl\}$ as
\[
\mu_1 = \frac{\theta_{1}\theta_{11}b(\theta_{111})-\theta_0\theta_{01}b(\theta_{011})}{\theta_{1}\theta_{11}-\theta_0\theta_{01}}, \ 
\mu_0 = \frac{(1-\theta_{1})\theta_{10}b(\theta_{101})-(1-\theta_0)\theta_{00}b(\theta_{001})}{(1-\theta_{1})\theta_{10}-(1-\theta_0)\theta_{00}},
\]
the causal paratemers can be estimated by plugging $\widehat\theta$ into $\mu_d$, denoted by $\widehat\mu_d$, for $d=0,1$. The principal average causal effect is estimated by $\widehat\tau=\widehat\mu_1-\widehat\mu_0$. The asymptotic properties of the plug-in estimators can be established by the sandwich estimator, as shown in the following theorem.

\begin{theorem} \label{prop1}
Under Assumptions \ref{ass:con}--\ref{ass:pi}, suppose there is no missingness. Let $\Gamma_d(\theta) = \partial\mu_d/\partial\theta$ for $d=0,1$, then
\begin{align*}
& n^{1/2}\left\{\widehat\mu_1-\mu_1\right\} \rightarrow_{d} N\left\{0, \ n\Gamma_1(\theta)^{\T}\cov(\widehat\theta)\Gamma_1(\theta)\right\}, \\
& n^{1/2}\left\{\widehat\mu_0-\mu_0\right\} \rightarrow_{d} N\left\{0, \ n\Gamma_0(\theta)^{\T}\cov(\widehat\theta)\Gamma_0(\theta)\right\}, \\
& n^{1/2}\left\{\widehat\tau-\tau\right\} \rightarrow_{d} N\left\{0, \ n\left(\Gamma_1(\theta)-\Gamma_0(\theta)\right)^{\T}\cov(\widehat\theta)\left(\Gamma_1(\theta)-\Gamma_0(\theta)\right)\right\}.
\end{align*}
\end{theorem}

When the outcome $Y$ is binary, the normal approximation may not have a satisfactory finite-sample performance, since the distribution of the estimator $\widehat\mu_d$ is skewed. To respect the fact that $\mu_d$ must lie in $[0,1]$, we may as well adopt $\logit(\mu_d)=\log\{\mu_d/(1-\mu_d)\}$ as the target estimand, which can range on the whole real line. The contrast $\tau^*=\logit(\mu_1)-\logit(\mu_0)$ reflects the log odds ratio between treatment and control for survived compliers. However, the log odds ratio does not enjoy collapsibility. The asymptotic properties for the maximized likelihood estimator of $\tau^*$ are provided in the Supplementary Material.
In the presence of missingness, analyses can be performed for each completed dataset, and inference is trivial to combine these results using Rubin's rule \citep{rubin2018multiple}.

\section{Simulation} \label{sec4}

In this section, we conduct a simulation study to assess the performance of the proposed method. Suppose there are $n \in \{500,2000,8000\}$ independent individuals in the sample. Each individual is assigned to the treated or control group with equal probability $r=0.5$. The compliance type is generated as
\begin{align*}
P(D(0)=1) = 0.3, \ P(D(1)=1 \mid D(0)=0) = 0.4.
\end{align*}
So the proportion of always-takers, compliers, and never-takers is 0.3, 0.28, and 0.42, respectively. The survival type is generated as
\begin{align*}
P(S(0)=1 \mid D(0),D(1)) &= 0.3 + 0.2 D(0) + 0.2 D(1), \\
P(S(1)=1 \mid D(0),D(1)) &= 0.3 + 0.3 D(0) + 0.3 D(1).
\end{align*}
The compliance type $(D(1),D(0))$ acts as the unmeasured confounder $U$, which can affect both $D$ and $S$.

To generate the outcome, we consider two settings. We first assume that the distribution of potential outcomes is homogeneous among compliance types (Case 1). That is, $Y(d)$ does not depend on $(D(1),D(0))$. As long as $S(d)=1$, we generate
\begin{align*}
\{Y(d) \mid S(d)=1\} \sim N(1+d, (0.8+0.2d)^2).
\end{align*}
Next, we consider another scenario where the distribution of potential outcomes is heterogeneous among compliance types (Case 2). For compliers,
\begin{align*}
\{Y(d) \mid G=c, S(d)=1\} \sim N(1+d, (0.8+0.2d)^2).
\end{align*}
For never-takers and always-takers,
\begin{align*}
\{Y(0) \mid G=n, S(0)=1\} &\sim N(1, 0.8^2) - N(0.5, 0.2^2), \\
\{Y(1) \mid G=a, S(1)=1\} &\sim N(2, 1.0^2) + N(0.3, 0.2^2).
\end{align*}
The observed variables under consistency are
\begin{align*}
D = ZD(1)+(1-Z)D(0), S = DS(1)+(1-D)S(0), Y = DY(1)+(1-D)Y(0).
\end{align*}
We assume there is no missingness. Assumptions \ref{ass:con}--\ref{ass:pi} hold under the data generating process.

To assess the robustness of the proposed method against principal ignorability (Assumption \ref{ass:pi}), which is the assumption most likely to be violated in practice, we further consider two scenarios. We assume that the potential outcome $Y(d)$ has cross-world reliance on $S(1-d)$. In Case 3, we add $S(1)$ to $Y(0)$ and add $0.5S(0)$ to $Y(1)$ under homogeneity (Case 1). In Case 4, we add $S(1)$ to $Y(0)$ and add $0.5S(0)$ to $Y(1)$ under heterogeneity (Case 2). The true treatment effect is empirically evaluated by the mean difference in $Y(1)$ and $Y(0)$ among survived compliers $\{D(1)>D(0), S(1)=S(0)=1\}$.

We compare two estimators. The first estimator is the proposed one (denoted by PACE), and the standard error of the treatment effect estimator can be calculated by Theorem \ref{prop1}. The second estimator is the two-stage least squares estimator (denoted by 2SLS), where we restrict the second-stage regression to the sample of observed survivors $\{S=1\}$. Table \ref{tab:simu} shows the bias, standard deviation (SD), standard error (SE), and coverage percentage (CP) of the nominal 95\% confidence interval in 10000 repeatedly generated datasets. The proposed method, PACE, has a negligible bias in Cases 1 and 2, where all assumptions hold. When principal ignorability is violated in Cases 3 and 4, the proposed method is biased, but the bias is smaller than that of 2SLS. The 2SLS estimator is biased in all settings because it cannot deal with the cross-world reliance of $S(d)$ on $(D(1),D(0))$. The confidence interval constructed by the asymptotic formula has a coverage percentage close to the nominal level. In all settings, the proposed method has a samller variance the 2SLS, indicating the efficiency of the proposed method. With the larger sample size, the standard deviation and standard error get smaller.

\begin{table}[!tbh]
\centering
\caption{The bias, standard deviation (SD), standard error (SE), and coverage percentage (CP) of the nominal 95\% confidence interval in the simulation study} \label{tab:simu}
\begin{tabular}{cccccccccc}
\toprule
 & & \multicolumn{4}{c}{Principal ignorability} & \multicolumn{4}{c}{Not principal ignorability} \\
\cmidrule(lr){3-6} \cmidrule(lr){7-10}
 & & \multicolumn{2}{c}{Homogeneous} & \multicolumn{2}{c}{Heterogeneous} & \multicolumn{2}{c}{Homogeneous} & \multicolumn{2}{c}{Heterogeneous} \\ 
 \cmidrule(lr){3-4} \cmidrule(lr){5-6} \cmidrule(lr){7-8} \cmidrule(lr){9-10}
 Size & Method & PACE & 2SLS & PACE & 2SLS & PACE & 2SLS & PACE & 2SLS \\
\midrule
 500 & Bias & 0.006 & 0.191 & -0.064 & 0.124 & 0.142 & 0.293 & 0.072 & 0.225 \\ 
 & SD & 0.394 & 0.475 & 0.497 & 0.600 & 0.438 & 0.498 & 0.572 & 0.628 \\ 
 & SE & 0.402 & 0.499 & 0.490 & 0.654 & 0.444 & 0.518 & 0.551 & 0.662 \\ 
 & CP & 0.969 & 0.948 & 0.966 & 0.963 & 0.948 & 0.925 & 0.933 & 0.944 \\ 
 2000 & Bias & -0.000 & 0.180 & -0.014 & 0.135 & 0.146 & 0.283 & 0.132 & 0.238 \\ 
 & SD & 0.177 & 0.224 & 0.215 & 0.287 & 0.197 & 0.239 & 0.242 & 0.302 \\ 
 & SE & 0.182 & 0.244 & 0.216 & 0.320 & 0.200 & 0.254 & 0.240 & 0.324 \\ 
 & CP & 0.958 & 0.905 & 0.954 & 0.947 & 0.885 & 0.812 & 0.890 & 0.895 \\ 
 8000 & Bias & 0.001 & 0.183 & -0.002 & 0.144 & 0.150 & 0.287 & 0.147 & 0.248 \\ 
 & SD & 0.088 & 0.114 & 0.106 & 0.145 & 0.098 & 0.120 & 0.119 & 0.151 \\ 
 & SE & 0.089 & 0.122 & 0.105 & 0.159 & 0.098 & 0.126 & 0.117 & 0.161 \\ 
 & CP & 0.948 & 0.689 & 0.948 & 0.875 & 0.658 & 0.369 & 0.733 & 0.668 \\
\bottomrule
\end{tabular}
\end{table}

In summary, the proposed method can consistently estimate the principal average causal effect. When the core identification assumption is violated, it is still more robust than the 2SLS.

\section{Application to Job Corps study} \label{sec5}

The U.S. Job Corps training program was a randomized education and job training program for disadvantaged youths. The study collected baseline covariates including gender, age, race, education status, and family information. There are 9240 complete cases, among which 5577 were assigned to Job Corps and 3663 were not assigned to Job Corps. In the first year after assignment, 84.6\% enrolled in education and/or vocational training in the treated group, and 50.6\% enrolled in education and/or vocational training in the control group. In the second year after the assignment, 52.3\% enrolled in education and/or vocational training in the treated group, and 42.0\% enrolled in education and/or vocational training in the control group. In our analysis, we regard the individual as receiving the training if he/she has ever enrolled in the training when analyzing the treatment effects later than the second year after assignment. So, 88.8\% in the treated group actually received the treatment, and 60.6\% in the control group actually received the treatment. We infer that the proportion of always-takers, compliers, and never-takers is 0.606, 0.282, and 0.112, respectively.

In one year after the assignment, 49.4\% have ever worked in the treated group, and 57.1\% have ever worked in the control group. In the second year after the assignment, the proportion of employment is 75.2\% in the treated group and 75.3\% in the control group. In the third year after assignment, the proportion of employment increases to 83.5\% in the treated group and 81.4\% in the control group. In the fourth year after the assignment, the proportion of employment increases to 83.7\% in the treated group and 81.3\% in the control group. We see that the employment status becomes stable in three years after the assignment. We take the logarithm of the weekly earnings. There is a trend for earnings to grow higher over time. Table \ref{tab:jc} lists the proportion of employment and average earnings in each level of the assigned treatment $Z$ and received treatment $D$.

\begin{table}[!tbh]
\centering
\caption{The proportion of employment and average earnings in each level of the assigned treatment $Z$ and received treatment $D$ in the Job Corps study} \label{tab:jc}
\begin{tabular}{ccccccccc}
\toprule
Year & \multicolumn{4}{c}{Employment $S$} & \multicolumn{4}{c}{Earnings $Y$ in logarithm} \\
\cmidrule(lr){2-5} \cmidrule(lr){6-9} 
$(Z,D)$ & (1,1) & (1,0) & (0,1) & (0,0) & (1,1) & (1,0) & (0,1) & (0,0) \\
\midrule
1 & 0.480 & 0.574 & 0.543 & 0.601 & 4.824 & 5.090 & 4.803 & 5.041 \\ 
2 & 0.749 & 0.772 & 0.745 & 0.766 & 4.723 & 4.841 & 4.661 & 4.868 \\ 
3 & 0.838 & 0.812 & 0.807 & 0.824 & 5.023 & 4.964 & 4.924 & 5.009 \\ 
4 & 0.840 & 0.815 & 0.827 & 0.792 & 5.207 & 5.128 & 5.160 & 5.168 \\
\bottomrule
\end{tabular}
\end{table}

\begin{figure}[!tbh]
\centering
\includegraphics[width=0.48\textwidth]{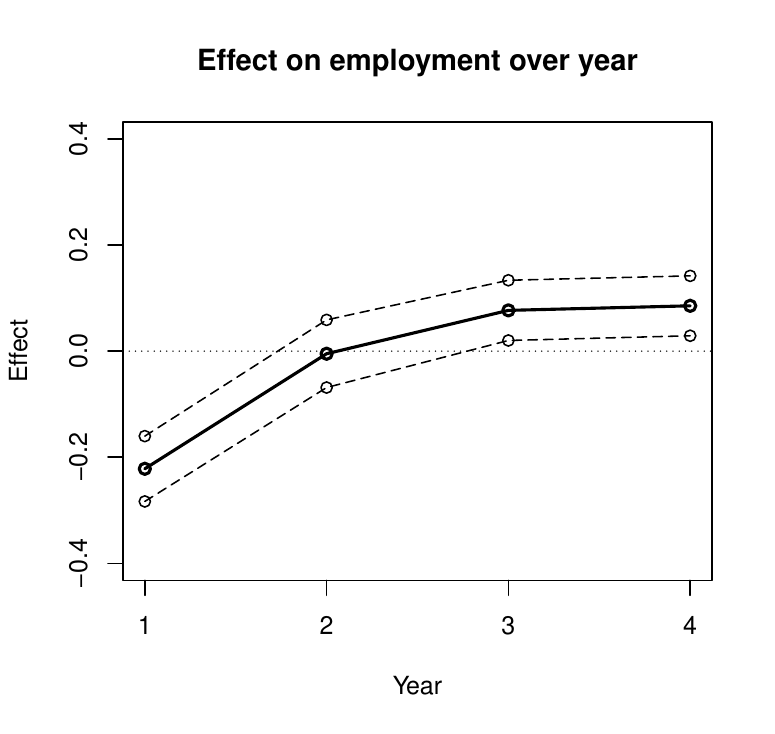}
\includegraphics[width=0.48\textwidth]{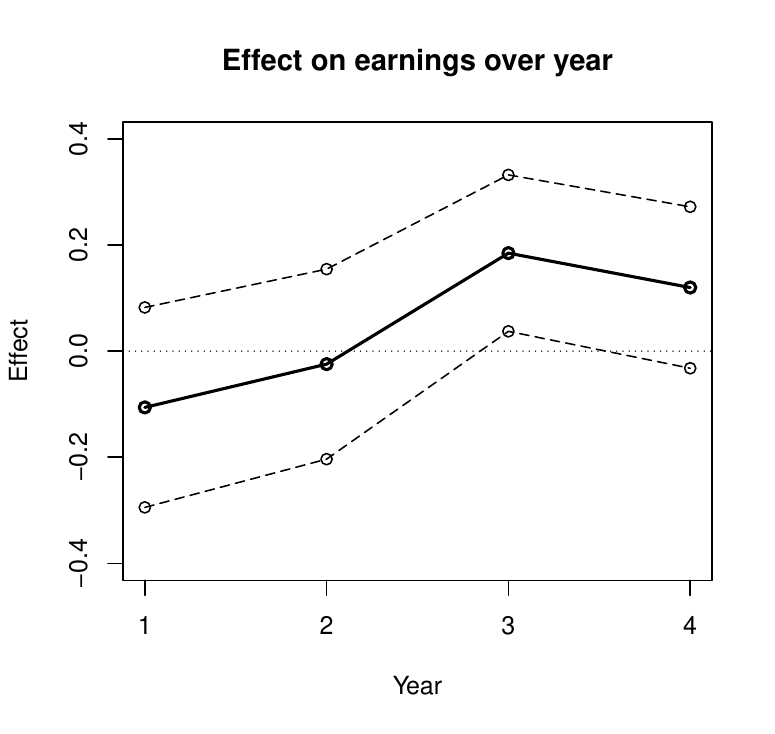}
\caption{Time-varying treatment effect on employment $E\{S(1)-S(0)\mid c\}$ and time-varying treatment effect on earnings $E\{Y(1)-Y(0) \mid cl\}$.} \label{fig:eff}
\end{figure}

We control six baseline covariates to model the earnings: gender, age, log household size, years of education, ever worked, and health. There are missing values in covariates. We perform 10 times of imputation. Finally, we use Rubin's rule to calculate the sampling variance. Figure \ref{fig:eff} shows the estimated treatment effect on employment $E\{S(1)-S(0) \mid c\}$ and the estimated treatment effect on weekly earnings $E\{Y(1)-Y(0) \mid cl\}$ with 95\% confidence intervals. In the first year after the assignment, there is a negative effect on employment. This is probably because the individuals spent their time on training, and they were not ready to enter the job market in the first year. Since the third year after the assignment, there has been a positive effect on employment, indicating that the training program helped the youths find jobs. The $P$-values are 0.0000, 0.8813, 0.0079, and 0.0030 for the four years, respectively.

The treatment effect on earnings is negative in the first year, indicating that joining the training program leads to lower earnings, probably because the individuals in the program have shorter work time. In the third year after the assignment, there is a significant and positive effect on earnings. The always-employed and compliers have 18.5\% higher salaries if they take the training program. The effect becomes slightly smaller in the fourth year because the weekly earnings without participating in the training program $\mu_0$ got higher. The $P$-values are 0.2700, 0.7885, 0.0140, and 0.1224 for these four years, respectively.

\begin{figure}[!tbh]
\centering
\includegraphics[width=0.48\textwidth]{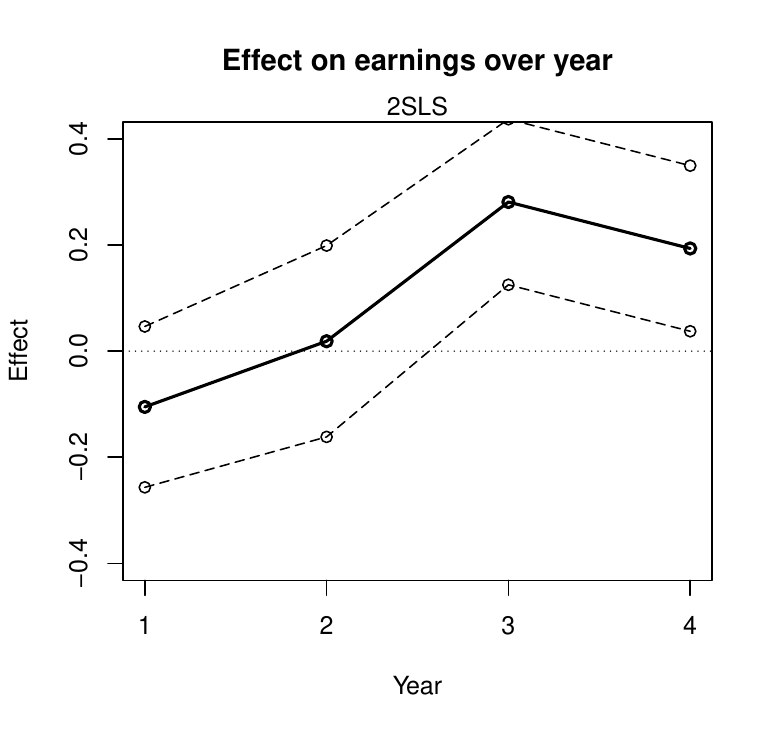}
\includegraphics[width=0.48\textwidth]{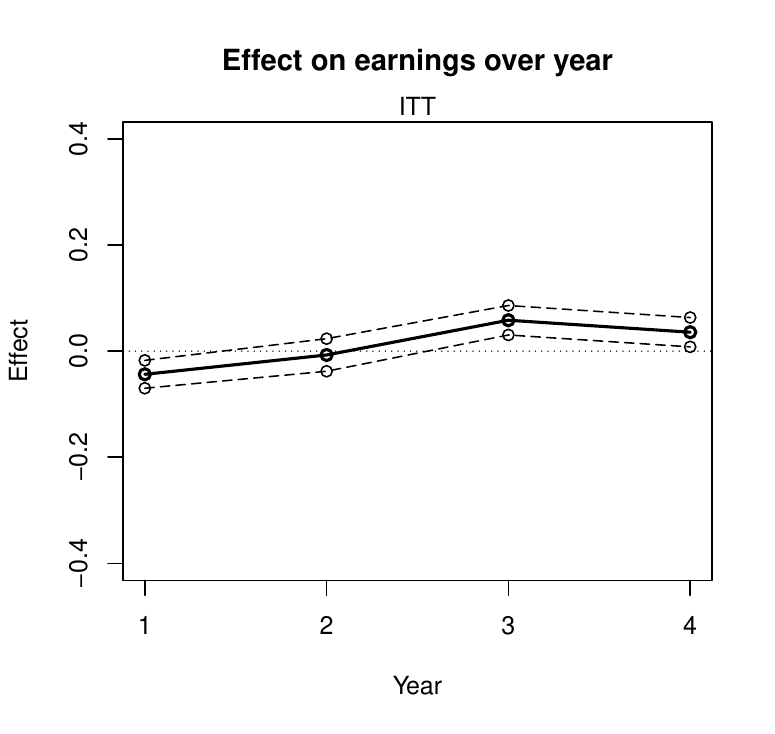} \\
\includegraphics[width=0.48\textwidth]{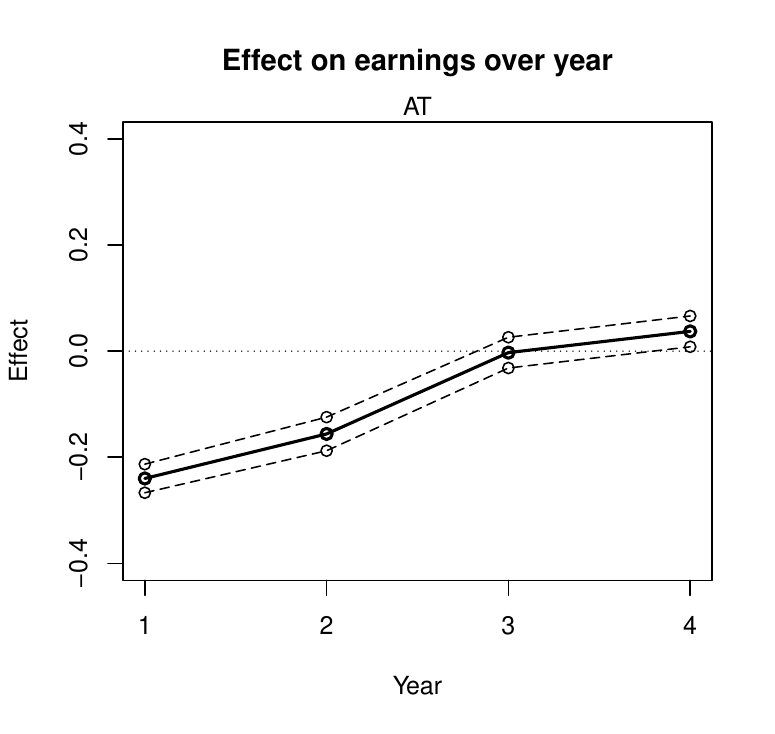}
\includegraphics[width=0.48\textwidth]{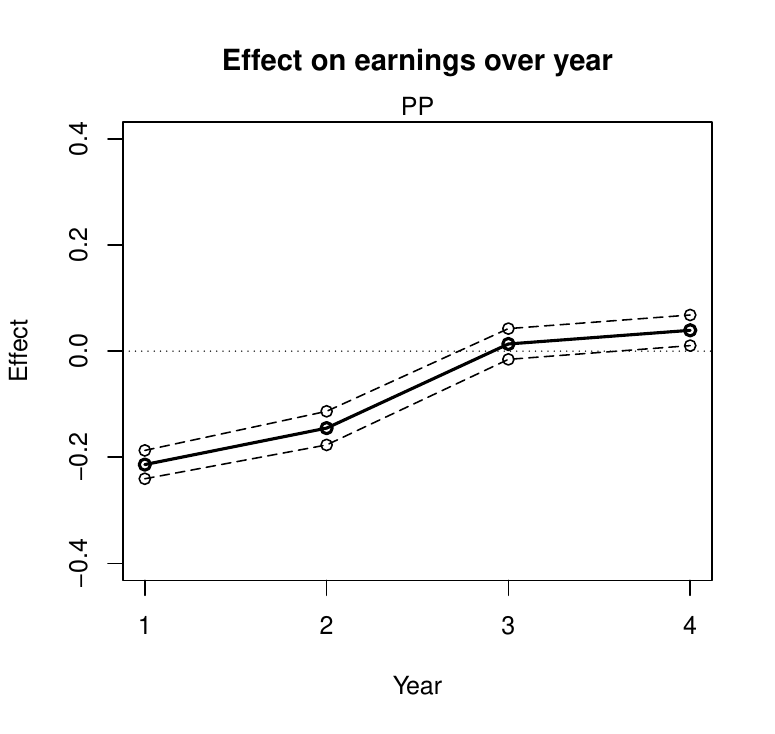}
\caption{Time-varying treatment effect on earnings $E\{Y(1)-Y(0) \mid cl\}$ estimated by 2SLS, ITT, AT and PP.} \label{fig:eff_n}
\end{figure}

Figure \ref{fig:eff_n} presents the estimated treatment effects by other methods. The point estimate of 2SLS is similar to that of the proposed method, indicating that the cross-world reliance of employment on the compliance type is weak. In the third and fourth years, ITT, AT, and PP underestimate the principal stratum treatment effect, indicating that compliers can benefit more from the training program. The confidence intervals given by ITT, AT, and PP are tighter because these methods do not account for the uncertainty in estimating the probability of survived (always-employed) compliers. 

The results of the proposed method have meaningful causal interpretations. However, since the target population, consisting of the survived compliers, is not individually identifiable, how to guide the results for future policy making is doubtful. Although we know the proportion of survived compliers, we do not know who will benefit from the training program. From the perspective of policymakers, the intention-to-treat analysis is the easiest to interpret, as it basically evaluates the effect of the treatment policy, regardless of whether the individuals adhere to the treatment assignment. In contrast, the proposed method evaluates the pure effect of the training program. The estimate by the proposed method is only valid for survived compliers, so we can never know how things would be if the training is made compulsory. Nevertheless, different methods approach the scientific question from different perspectives, and assessment by different methods provides a comprehensive knowledge of how the treatment can make an effect. In the Supplementary Material, we relax the identification assumptions by adjusting for baseline covariates and infer an interventionist complier average treatment effect. We find that the treatment effect on earnings is significant for the third and fourth years in compliers.

\section{Discussion} \label{sec6}

Post-treatment events bring challenges to conducting causal analyses of the treatment effect. Although the baseline characteristics are naturally balanced by randomization, these characteristics are not balanced anymore if there is post-treatment selection. In this paper, we consider the broken randomized experiments with non-compliance, truncation-by-death, and missing data. Non-compliance leads the received treatment to differ from the assigned treatment, and the outcomes after the received treatment are still meaningful, reflecting a mixed effect of the treatment and possibly unmeasured compliance type. Truncation-by-death renders the outcomes ill-defined or undefined. The survival type also embeds information about the underlying characteristics of individuals. When there are non-compliance, truncation-by-death, and missingness simultaneously in a randomized trial, we propose a framework to identify and estimate the principal stratum average treatment effect. The causal effect is defined in a subpopulation called the survived compliers. Although we cannot identify who belongs to this subpopulation or the proportion of this subpopulation, we can still identify the treatment effect in this subpopulation. 

There are some limitations for the estimand. First, the identification involves untestable assumptions. The four assumptions on non-compliance are in line with the assumptions to identify the complier average causal effect (CACE) or the local average treatment effect (LATE). Many efforts have been devoted to finding testable implications for these assumptions \citep{yang2014dissonant, mogstad2018using, kedagni2020generalized}. The principal ignorability assumption is hard to justify, and sensitivity analysis may be helpful to assess the robustness of substantial results to the violation of principal ignorability. A conservative approach is to find bounds for the treatment effect \citep{long2013sharpening, marden2018implementation}. Whether the bounds are informative is worthy of discussion. Utilizing additional information such as substitutional variables or post-treatment correlates, principal ignorability can be relaxed \citep{tchetgen2014identification, wang2017identification}.

Second, we did not consider adjustments of measured covariates, especially in the identification assumptions. Covariate adjustment allows for studying causal effects in observational studies and fosters more efficient estimation \citep{feller2017principal, jiang2022multiply}. However, relaxing ignorability assumptions by conditioning on covariates could incur other challenges because the average potential outcomes need to be integrated over the distribution of covariates in the target principal stratum, while the proportion of this principal stratum given covariates has not been identified. In Supplementary Material B, we provide some identification results with baseline covariates, under an additional survival monotonicity assumption. This issue is crucial if we aim to extend the proposed framework to observational studies. In Supplementary Material C, we propose an interventionist estimand on compliers, which does not need survival monotonicity.

Third, as a common criticism of principal stratification, it is hard to interpret the causal effect and generalize the result to a new population \citep{pearl2011principal, vanderweele2011principal}. The causal effect is defined on a latent subpopulation where both potential outcomes theoretically exist. An individual in this subpopulation must comply with treatment assignments and be destined to have an outcome. Even if a policy is effective, policymakers cannot easily implement it because they never know who will follow the policy instructions and benefit from the policy. As a tool, principal stratification defines the estimand in a subpopulation where all potential outcomes are well-defined and comparable. As a goal, the principal stratum should be consistent with the scientific interest. The application scenario should be carefully justified so that the estimated treatment effect can be meaningfully interpreted.

\section*{Acknowledgements}

We thank the associate editor and reviewers for theor comments. We thank the support from the Save 2050 Programme jointly sponsored by the Swarma Club and X-Order.
JEL: C18, C39, C51

\section*{Funding information}

This work is supported by the National Social Science Fund of China, Grant No.~19BJY108.

\section*{Data availability statement} 

The data we use in our work is publicly available in the R package \texttt{causalweight}. 

\section*{Supplementary Material}

The supplementary material includes estimands for binary outcomes and estimands under assumptions adjusting for baseline covariates.

\bibliographystyle{apalike}
\bibliography{ref}



\section*{Supplementary material}

# Appendix: Identification of the causal effect with covariates

For simplicity, suppose there are no missing values, otherwise multiple imputation approaches can be applied before data analysis. Here we restate the identification assumptions. The independencies in Assumptions 1–6 should be conditioned additionally on baseline covariates $X$, as listed below.

For notation simplicity, we assume there is no missingness (otherwise we just need to add the missingness indicator in the conditional expectations). In the presence of covariates, one can easily identify and estimate the conditional principal average causal effect (or called the local covariate-specific average treatment effect),

$$E\{Y(1) \mid G = cl, X\} = \frac{E\{DSY \mid Z = 1, X\} - E\{DSY \mid Z = 0, X\}}{E\{DS \mid Z = 1, X\} - E\{DS \mid Z = 0, X\}},$$
$$E\{Y(0) \mid G = cl, X\} = \frac{E\{(1-D)SY \mid Z = 1, X\} - E\{(1-D)SY \mid Z = 0, X\}}{E\{(1-D)S \mid Z = 1, X\} - E\{(1-D)S \mid Z = 0, X\}}.$$

However, when it comes to the population-level average causal effect, the main obstacle for identification is that the proportion of the $cl$ stratum is not identifiable. Indeed, solely with principal ignorability, one can never distinguish $cl$ and $cp$ strata among the treated, nor

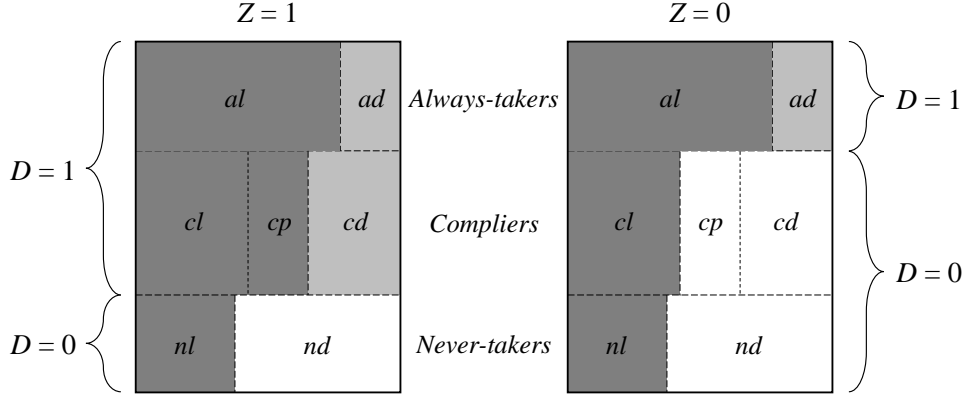

Figure 4: Monotonicity on survival. Dark shaded are the observed survivors.

distinguish $cl$ and $ch$ strata among the controlled using observed data. Therefore, additional assumptions should be made to identify the population-level estimand.

**Assumption 7 (Monotonicity on survival)** $S(1) \geq S(0)$ *almost surely.*

Under monotonicity on survival, the proportions of each principal stratum are all identifiable. Figure 4 illustrates the constitution of observed data in each group. Particularly, the proportion of the $cl$ stratum conditional on covariates is identified as

$$P(G = cl \mid X = x) = P(D = 0 \mid Z = 0)E\{S \mid (1-Z)(1-D)\Delta_S = 1, X = x\}$$
$$- P(D = 0 \mid Z = 1)E\{S \mid Z(1-D)\Delta_S = 1, X = x\}.$$

The first term is the conditional proportion of $cl$ and $nl$, and the second term is the conditional proportion of $nl$. Next, the population-level principal average causal effect can be obtained by taking expectation for the conditional principal average causal effect over the covariate distribution on the $cl$ stratum.

**Theorem 4** *Under the conditional version of Assumptions 1–6 and Assumption 7, the principal average causal effect $\tau$ is nonparametrically identifiable,*

$$E\{Y(1) \mid G = cl\} = \frac{\int_{\mathcal{X}} E\{Y(1) \mid G = cl, X = x\} P(G = cl \mid X = x) dP(x)}{\int_{\mathcal{X}} p(G = cl \mid X = x) dP(x)},$$

$$E\{Y(0) \mid G = cl\} = \frac{\int_{\mathcal{X}} E\{Y(0) \mid G = cl, X = x\} P(G = cl \mid X = x) dP(x)}{\int_{\mathcal{X}} p(G = cl \mid X = x) dP(x)},$$

*where $P(x)$ is the distribution function of $X$ and $\mathcal{X}$ is the support of $X$.*

The identification formula motivates a regression estimator for the target estimand. In fact, one can construct weighting estimators using the propensity score and principal score (Feller et al., 2017; Ding and Lu, 2017). Then multiply robust estimator can be derived by finding the efficient influence function of the estimand (Wang and Tchetgen Tchetgen, 2018; Jiang et al., 2022). When models are correctly specified, the efficient influence function based estimator may enjoy better efficiency than the regression estimator and weighting estimator.

Monotonicity is untestable. Whether monotonicity on survival is reasonable depends on specific scientific questions. There are some variants of the individual-level monotonicity such as stochastic monotonicity (Lee et al., 2010), which assumes a known relation, also relying untestable prior knowledge, between the potential survivals. To relax the principal ignorability assumption, we can invoke a substitutional (negative control) variable related with potential survivals but not directly related with potential outcomes (Wang et al., 2017; Deng et al., 2021). The substitutional variable embeds information of the principal stratum, so the different distribution of mean outcomes can be separated by recovering the information of $G$ using $V$. The data structure would be more complex and may not be easily interpretable for actual scenarios.



\end{document}